\documentclass[prb,aps,preprint,eqsecnum]{revtex4}
\usepackage[centertags]{amsmath}
\usepackage{amssymb}
\usepackage{amsthm}
\usepackage{graphicx}
\usepackage{epsfig}

\begin{document}

\title{Quantum state-dependent diffusion and multiplicative noise:
a microscopic approach}
\author{Debashis~Barik and Deb~Shankar~Ray{\footnote{
Email address: pcdsr@mahendra.iacs.res.in}}} \affiliation{Indian
Association for the Cultivation of Science, Jadavpur, Kolkata 700
032, India}

\begin{abstract}
The state-dependent diffusion, which concerns the Brownian motion
of a particle in inhomogeneous media has been described
phenomenologically in a number of ways. Based on a
system-reservoir nonlinear coupling model we present a microscopic
approach to quantum state-dependent diffusion and multiplicative
noise in terms of a quantum Markovian Langevin description and an
associated Fokker-Planck equation in position space in the
overdamped limit. We examine the thermodynamic consistency and
explore the possibility of observing a quantum current, a generic
quantum effect, as a consequence of this state-dependent
diffusion similar to one proposed by B\"{u}ttiker [Z. Phys. B
{\bf 68}, 161 (1987)] in a classical context several years ago.

\end{abstract}

\maketitle

\section{Introduction}

Almost three decades ago Landauer \cite{lan1,lan2} explored the
problem of characterizing nonequilibrium steady states  in the
transition kinetics between the two locally stable states in
bistable systems. His main idea was that the relative stability
of a particle diffusing in a bistable potential can be altered by
an intervening hot layer which has the effect of pumping particles
from a globally stable region to a metastable region. No detailed
consideration of immediate neighbourhood of the two states is
important. In formulating the problem in terms of diffusion
equation it was realized that one needs state dependence of
diffusion for a correct description of the effect and more
generally a careful analysis of the problem of diffusion in
inhomogeneous media in a wider context was necessary. This was
carried out by van Kampen \cite{van} and others \cite{lin} in
eighties. An important consequence of state-dependent diffusion
or noise as suggested by B\"{u}ttiker \cite{but} is the
generation of current, in absence of any externally applied
fields, which occurs in presence of periodic diffusion of a
particle in a spatially periodic potential with same periodicity
but differing in phase. This rectification of state dependent
noise resulting in a directed transport and state-dependent
diffusion play important role in several areas of condensed
matter physics on the mesoscopic scale \cite{kuz,hak,mar,baut1}
and furthermore in ratchet problems \cite{mag,jul,rei,kla} in a
wider perspective.

The physics of state-dependent diffusion can be described
phenomenologically in a number of ways. As reported in the
literature \cite{van,sol} the diffusion term for Brownian particle
may assume several forms, notably $\frac{\partial}{\partial
q}D(q)\frac{\partial}{\partial q} P(q,t)$ or
$\frac{\partial^2}{\partial q^2}D(q)P(q,t)$ or the other like
$\frac{\partial}{\partial q}D\frac{\partial P(q,t)}{\partial q}$
supplemented by 'thermal potential' or state-dependent drift term.
Here $D(q)$ refers to diffusion coefficient and $P(q,t)$ is the
probability distribution function for the particle. Two important
points are now noteworthy. First, the phenomenological forms of
diffusion coefficient being different, it is easy to realize that
they do not have a common microscopic Hamiltonian origin. Thus the
physics of diffusion in inhomogeneous media is somewhat
model-dependent \cite{sol} and the search for a "correct" form of
diffusion remains a debatable issue. Second, the diverse forms
notwithstanding, the generalization of Boltzmann factor,
$\exp(-V(q)/k_B T)$ (which governs the system at thermal
equilibrium) for state-dependent diffusion in the steady state
assumes a common structure,

\begin{equation}\label{1.1}
P_{st}(q)\sim \exp[-\phi(q)]
\end{equation}

with $\phi(q)=\int_0^q \frac{V^\prime(q^\prime)}{D(q^\prime)}
\;dq^\prime$, $V(q)$ being the potential field. The steady state
distribution (\ref{1.1}) implies that the effective potential
$\phi(q)$ is nonlocal in space. The generality in the structure
of $\phi(q)$ is such that it may include the spatial variation of
temperature, diffusion or drift coefficient as specific cases as
considered separately by several authors \cite{van,but,sol}. In
the Langevin scheme of description, on the other hand
state-dependent diffusion has received attention under
"multiplicative noise"
\cite{san,san1,jay,san2,mas,lin1,sak,tai,san3}. The microscopic
origin of multiplicative noise within the framework of standard
paradigm of system-reservoir Hamiltonian that includes a variety
of model calculations is the nonlinear coupling between the
system and bath coordinates which leads to nonlinear dissipation.
A thermodynamically consistent approach in this context was put
forward in early eighties by Lindenberg and coworkers \cite{lin}.
An exact Fokker-Planck equation for time and space dependent
friction was derived by Pollak \textit{et al} \cite{pol} several
years ago. Along with these formal developments
\cite{sol,lin,pol,san,san1,jay,san2,mas,lin1,sak,tai,san3}, the
theories of multiplicative noise have found wide applications in
several areas, \textit{e.g.}, activated processes \cite{rate},
stochastic resonance \cite{stka}, laser and optics \cite{les},
signal processing \cite{sig}, fluctuation-induced transport
\cite{baut,tran}, noise-induced transitions \cite{tran1} etc.

In this paper we address the problem of Langevin equation with
multiplicative noise and state-dependent diffusion for a
thermodynamically closed system in a \textit{quantum mechanical}
context. Although the quantum-mechanical system-reservoir linear
coupling model for microscopic description of additive noise and
linear dissipation which are related by fluctuation-dissipation
relation is well-known over many decades in several fields
\cite{wei,lui}, the nature of nonlinear coupling and its
consequences have been explored with renewed interest only
recently. For example, it has been observed that the quantum
dissipation can reduce the appearance of metastable state and
barrier drift in a double well potential \cite{bao}. Tanimura and
co-workers \cite{tani} have extensively used nonlinear coupling
in modeling elastic and inelastic relaxation mechanisms and their
interplay in vibrational and Raman spectroscopy. The role of
inhomogeneous dissipation in reducing quantum decay rate has also
been explored very recently \cite{bao1}. Based on a coherent
state representation of quantum noise operator and Wigner
canonical thermal distribution for harmonic bath oscillators
\cite{hil} nonlinearly coupled to a system we would like to
develop a microscopic approach to quantum state-dependent
diffusion and quantum multiplicative noise. Specifically, our aim
is:

(i) to derive a quantum Smoluchowski equation for
state-dependent diffusion on the basis of a system-reservoir
model and analyze the role of nonlinear coupling in
state-dependence of quantum diffusion and dissipation.

(ii) to seek for a correspondence between this quantum equation and
its phenomenological counterpart.

(iii) to derive a quantum generalization of Boltzmann factor for
state-dependent diffusion in the steady state
(\ref{1.1}) and check its thermodynamic consistency.

(iv) to explore the possibility of observing a directed transport
as a consequence of state dependent quantum diffusion in the
spirit of B\"{u}ttiker as an immediate application.

The outlay of the paper is as follows: In Sec.II we develop the
scheme of quantum Brownian motion for multiplicative noise on the
basis of a nonlinearly coupled system-reservoir model of
Zwanzig-type form within a Markovian description. Sec.III is
devoted to quantum Smoluchowski equation for
state-dependent diffusion which corresponds to a specific form of
a classical phenomenological equation. We derive a quantum
mechanical generalization of Boltzmann factor for state-dependent
diffusion. In Sec.IV we carry out an application to a system with
spatially periodic potential and periodic diffusion function with
same periodicity but differing in phase to demonstrate a nonzero
current. The paper is concluded in Sec.V.

\section{Quantum multiplicative noise}

\subsection{General aspects; Langevin equation}

We consider a particle of unit mass coupled to a medium comprised
of a set of harmonic oscillators with frequency $\omega_j$. This
is described by the following system-bath Hamiltonian
\cite{bao,bao1,tani}.

\begin{equation}\label{2.1}
\hat{H}=\frac{\hat{p}^2}{2}+V(\hat{q})+\sum_j \left[\frac{\hat{p}^2_j}{2}+
\frac{1}{2}\left(\omega_j \hat{x}_j - \frac{c_j}{\omega_j} f(\hat{q})\right)^2\right]
\end{equation}

Here $\hat{q}$ and $\hat{p}$ are the coordinate and momentum
operators of the particle and the $\{\hat{x}_j,\hat{p}_j\}$ are
the set of coordinate and momentum operators for the bath
oscillators with unit mass. The system particle is coupled to the
bath oscillators nonlinearly through the general coupling terms
$\frac{c_j}{\omega_j}f(\hat{q})$. $c_j$ is the coupling strength.
The Hamiltonian Eq.(\ref{2.1}) is different from Zwanzig
\cite{zwa} form of system-bath Hamiltonian where the coupling is
linear with respect to system coordinate. The classical
counterpart \cite{lin} of the form (\ref{2.1}) is known for more than two
decades and also the nonlinear coupling in quantum system has
been studied in several contexts \cite{bao,tani}. The potential
$V(\hat{q})$ is due to external force field for the system
particle. The coordinate and momentum operators follow the usual
commutation relations $[\hat{q}, \hat{p}]=i \hbar$ and
$[\hat{x}_j, \hat{p}_k]=i \hbar \delta_{jk}$. The presence of
counter term in the Hamiltonian ensures that the potential
$V(\hat{q})$ felt by the particle does not get modified due to
heat bath.

We now use Eq.(\ref{2.1}) to obtain the following dynamical
equations for the position and momentum operators:

\begin{equation}\label{2.2}
\dot{\hat{q}}=\frac{\partial \hat{H}}{\partial \hat{p}}=\hat{p}
\end{equation}

\begin{equation}\label{2.3}
\dot{\hat{p}}=-\;\frac{\partial \hat{H}}{\partial
\hat{q}}=-V^\prime (\hat{q}) +f^\prime
(\hat{q})\sum_j\frac{c_j}{\omega_j}\left(\omega_j \hat{x}_j -
\frac{c_j}{\omega_j} f(\hat{q})\right)
\end{equation}

where the dot(.) indicates derivative with respect to time and
the prime ($\prime$) refers to derivative with respect to
$\hat{q}$.

Similarly we have the dynamical equations of motion for the bath
oscillators $(j=1,2,3...)$

\begin{equation}\label{2.4}
\dot{\hat{x}}_j=\frac{\partial \hat{H}}{\partial \hat{p}_j}=\hat{p}_j
\end{equation}

\begin{equation}\label{2.5}
\dot{\hat{p}}_j=-\;\frac{\partial \hat{H}}{\partial
\hat{x}_j}=-\;\omega_j^2\hat{x}_j
+c_j f(\hat{q})
\end{equation}

To eliminate the bath degrees of freedom from the equations of
motion of the system we first obtain a solution for the
position operator $\hat{x}_j$ by formally solving the above
equations and then make use of the solution in Eq.(\ref{2.3})
followed by some rearrangement. This yields the generalized operator
Langevin equation for the system particle.

\begin{eqnarray}
\dot{\hat{q}}(t)&=&\hat{p}(t)\label{2.6}\\
\dot{\hat{p}}(t)&=&-V^\prime (\hat{q}(t))-f^\prime
(\hat{q}(t))\int_0^t f^\prime (\hat{q}(t^\prime))
\gamma(t-t^\prime)\hat{p}(t^\prime)
dt^\prime+f^\prime(\hat{q}(t)) \hat{\eta}(t)\label{2.7}
\end{eqnarray}

where the noise operator $\hat{\eta}(t)$ and the memory kernel
$\gamma(t)$ are given by

\begin{equation}\label{2.8}
\hat{\eta} (t) = \sum_j \left [ \left \{
\frac{\omega_j^2}{c_j}\;\hat{x}_j (0) - f(\hat{q}(0)) \right \}
\frac{c_j^2}{\omega_j^2} \cos \omega_j t +
\frac{c_j}{\omega_j} \hat{p}_j (0) \sin \omega_j t \right]
\end{equation}

and

\begin{equation}\label{2.9}
\gamma(t)=\sum_j \frac{c_j^2}{\omega_j^2} \;\cos\omega_j t
\end{equation}

It is clear from the operator Langevin equation Eq.(\ref{2.7})
for the system that the noise operator is multiplicative and the
dissipative term is nonlinear with respect to system coordinate
due to the nonlinear coupling term in the system-bath
Hamiltonian. In the case of linear coupling, \textit{i.e.},
$f(\hat{q})=\hat{q}$ the Eq.(\ref{2.7}) reduces to a quantum
generalized Langevin equation \cite{wei} in which the noise term
is additive and the dissipative term is linear.

Since the system is thermodynamically closed, \textit{i.e.}, the
fluctuation and the dissipation originates from the same origin,
the detailed balance condition must be satisfied. The noise properties
of $\hat{\eta}(t)$ can be derived by using suitable canonical thermal
distribution of bath coordinates and momenta operators at $t=0$
to obtain;

\begin{equation}\label{2.10}
\langle \hat{\eta}(t)\rangle_{QS}=0
\end{equation}

\begin{equation}\label{2.11}
\frac{1}{2}\langle\hat{\eta}(t)\hat{\eta}(t^\prime)+\hat{\eta}(t^\prime)\hat{\eta}(t)
\rangle_{QS}=\frac{1}{2}\sum_j \frac{c_j^2}{\omega_j^2}\; \hbar
\omega_j \left(\coth \frac{\hbar \omega_j}{2 k_B T}\right) \cos
\omega_j (t-t^\prime)
\end{equation}

Here $\langle....\rangle_{QS}$ implies quantum statistical
average on bath degrees of freedom and is defined as

\begin{equation}\label{2.12}
\langle \mathcal{\hat{O}}
\rangle_{QS}=\frac{Tr\mathcal{\hat{O}}\exp(-\hat{H}_{bath}/k_B T
)}{Tr\exp(-\hat{H}_{bath}/k_B T )}
\end{equation}

for any bath operator
$\mathcal{\hat{O}}\left(\{\frac{\omega_j^2}{c_j}
\hat{x}_j-f(\hat{q})\}, \{\hat{p}_j\}\right)$, where
$\hat{H}_{bath}=\sum_j\left[\frac{\hat{p}_j^2}{2}+\frac{1}{2}(\omega_j
\hat{x}_j -\frac{c_j}{\omega_j}f(\hat{q}))^2\right]$ at $t=0$. By
trace we mean the usual quantum statistical average.
Eq.(\ref{2.11}) is the fluctuation-dissipation relation expressed
in terms of noise operators appropriately ordered in the quantum
mechanical sense.

In the Markovian limit the generalized quantum Langevin equation
Eq.(\ref{2.7}) reduces to the form

\begin{subequations}
\begin{eqnarray}
\dot{\hat{q}}(t)&=&\hat{p}(t)\label{2.13a}\\
\dot{\hat{p}}(t)&=&-V^\prime(\hat{q}(t))-\Gamma\;
[f^\prime(\hat{q}(t))]^2\; \hat{p}(t)+f^\prime(\hat{q}(t))
\hat{\eta}(t)\label{2.13b}
\end{eqnarray}
\end{subequations}

where $\Gamma$ is dissipation constant in the Markovian limit.

To construct a c-number quantum Langevin equation we proceed
\cite{skb,db1,db2,bk1,bk2} as follows. We carry out a quantum
mechanical average of Eq.(\ref{2.13a}) and Eq.(\ref{2.13b}) to get

\begin{eqnarray}
\dot{q}&=&p\label{2.14}\\
\dot{p}&=&-\langle V^\prime(\hat{q}) \rangle-\Gamma \langle
[f^\prime(\hat{q})]^2 \hat{p}\rangle+\langle f^\prime(\hat{q})
\hat{\eta}(t)\rangle\label{2.15}
\end{eqnarray}

where $q=\langle \hat{q} \rangle$ and $p=\langle \hat{p}
\rangle$. The quantum mechanical average $\langle ... \rangle$ is
taken over the initial product separable quantum states of the
particle and the bath oscillators at $t=0$,
$|\phi\rangle\{|\alpha_1\rangle|\alpha_2\rangle....|\alpha_N\rangle\}$.
Here $|\phi\rangle$ denotes any arbitrary initial state of the
system and $|\alpha_j\rangle$ corresponds to the initial coherent
state of the j-th bath oscillator. $|\alpha_j\rangle$ is given by
$|\alpha_j\rangle=\exp(-|\alpha_j^2|/2)\sum_{n_j=0}^\infty
(\alpha_j^{n_j}/\sqrt{n_j!})|n_j\rangle$, $\alpha_j$ being
expressed in terms of the mean values of the shifted coordinate
and momentum of the j-th oscillator,
$\{\frac{\omega_j^2}{c_j}\langle \hat{x}_j(0)\rangle-\langle
f(\hat{q}(0))
\rangle\}=\sqrt{\frac{\hbar}{2\omega_j}}(\alpha_j+\alpha_j^*)$
and $\langle \hat{p}_j(0)\rangle=\sqrt{\frac{\hbar \omega_j
}{2}}(\alpha_j^*-\alpha_j)$, respectively.

Since the quantum mechanical average is taken over the initial
product separable quantum states of the particle and the bath
oscillators the Eq.(\ref{2.14}-\ref{2.15}) can be written as:

\begin{subequations}
\begin{eqnarray}
\dot{q}&=&p\label{2.16a}\\
\dot{p}&=&-\langle V^\prime(\hat{q}) \rangle-\Gamma \langle
[f^\prime(\hat{q})]^2 \hat{p}\rangle+\langle
f^\prime(\hat{q})\rangle \eta(t)\label{2.16b}
\end{eqnarray}
\end{subequations}

where $\eta(t)=\langle\hat{\eta}(t)\rangle$, $\eta(t)$ is now a
classical-like noise term, which, in general, is a non-zero
number because of the quantum mechanical averaging and is given by

\begin{equation}\label{2.17}
\eta (t) = \sum_j \left [ \left \{
\frac{\omega_j^2}{c_j}\;\langle\hat{x}_j (0)\rangle - \langle
f(\hat{q}(0))\rangle \right \} \frac{c_j^2}{\omega_j^2} \cos
\omega_j t + \frac{c_j}{\omega_j}\; \langle \hat{p}_j (0) \rangle
\sin \omega_j t \right]
\end{equation}

To realize $\eta(t)$ as an effective c-number noise we now
introduce the ansatz \cite{hil,skb,db1,db2,bk1,bk2} that the
momentum $\langle \hat{p}_j (0) \rangle$ and the shifted
coordinates $\{\frac{\omega_j^2}{c_j}\;\langle\hat{x}_j
(0)\rangle - \langle f(\hat{q}(0))\rangle \}$ of the bath
oscillators are distributed according to a canonical distribution
of Gaussian form as:

\begin{eqnarray}
&\mathcal{P}_j \left(\{\frac{\omega_j^2}{c_j}\;\langle\hat{x}_j
(0)\rangle - \langle f(\hat{q}(0))\rangle \},\langle \hat{p}_j
(0) \rangle\right)\nonumber\\
&=\mathcal{N} \exp\left\{-\;\frac{[\;\langle \hat{p}_j
(0)\rangle^2+\frac{c_j^2}{\omega_j^2}\{\frac{\omega_j^2}{c_j}\;\langle\hat{x}_j
(0)\rangle - \langle f(\hat{q}(0))\rangle \}^2]}{2 \hbar \omega_j
\left(\bar{n}_j(\omega_j)+\frac{1}{2}\right) }\right\}\label{2.18}
\end{eqnarray}

so that for any quantum mechanical mean value,
$\mathcal{O}_j\left(\{\frac{\omega_j^2}{c_j}\;\langle\hat{x}_j
(0)\rangle - \langle f(\hat{q}(0))\rangle \},\langle \hat{p}_j
(0)\rangle\right)$ of the bath operator, its statistical average
$\langle...\rangle_S$ is

\begin{equation}\label{2.19}
\langle \mathcal{O}_j\rangle_S=\int \mathcal{O}_j \mathcal{P}_j\;
d\langle \hat{p}_j (0)\rangle\;
d\{\frac{\omega_j^2}{c_j}\langle\hat{x}_j (0)\rangle - \langle
f(\hat{q}(0))\rangle \}
\end{equation}

Here $\bar{n}_j(\omega_j)$ indicates the average thermal photon
number of the j-th oscillator at the temperature $T$ and
$\bar{n}_j(\omega_j)=[\exp(\frac{\hbar \omega_j}{ k_B T})-1]^{-1}$
and $\mathcal{N}$ is the normalization constant.

The distribution $\mathcal{P}_j$ (Eq.(\ref{2.18})) and the
definition of statistical average Eq.(\ref{2.19}) imply that
c-number noise $\eta(t)$ must satisfy

\begin{equation}\label{2.20}
\langle \eta(t) \rangle_S=0
\end{equation}

\begin{equation}\label{2.21}
\langle \eta(t) \eta(t^\prime) \rangle_S=\frac{1}{2}\sum_j
\frac{c_j^2}{\omega_j^2}\; \hbar \omega_j \left(\coth \frac{\hbar
\omega_j}{2 k_B T}\right) \cos \omega_j (t-t^\prime)
\end{equation}

which are equivalent to (\ref{2.10}) and (\ref{2.11}), respectively.

In the Markovian limit the noise correlation becomes

\begin{subequations}
\begin{eqnarray}
\langle \eta(t) \eta(t^\prime) \rangle_S&=&2D_0\delta(t-t^\prime)\label{2.22a}\\
D_0&=&\frac{1}{2}\Gamma \hbar \omega_0
\left(\bar{n}(\omega_0)+\frac{1}{2}\right)\label{2.22b}
\end{eqnarray}
\end{subequations}

where $\omega_0$ is the average bath frequency and the spectral
density function is considered in the Ohmic limit.

The Eqs.(\ref{2.20}-\ref{2.21}) imply that the c-number noise
$\eta(t)$ is such that it is zero centered and satisfies the
standard fluctuation-dissipation relation as expressed in
Eq.(\ref{2.11}). It is easy to recognize that the ansatz
(\ref{2.18}) is a canonical thermal Wigner distribution function
for a shifted harmonic oscillator (obtained as an exact solution
of Wigner equation \cite{hil} for harmonic oscillator) which
always remains a positive definite function. A special advantage
of using this distribution function is that it remains valid as a
pure state nonsingular distribution function even at $T=0$. At the
same time the distribution of quantum mechanical mean values of
the bath oscillators reduces to classical Maxwell-Boltzmann
distribution in the thermal limit $\hbar \omega \ll k_B T$.
Furthermore this procedure allows us to bypass operator ordering
prescription of Eq.(\ref{2.11}) for deriving noise properties of
the bath oscillators and to identify $\eta(t)$ as a classical
looking noise with quantum mechanical content. We also mention
that instead of Wigner function it is also possible to employ
Glauber-Sudarshan distribution function to derive quantum
fluctuation-dissipation relation \cite{skb,db1,db2,bk1,bk2}.

We now return to Eq.(\ref{2.16a}-\ref{2.16b}) to add
$V^\prime(q)$, $\Gamma[f^\prime(q)]^2p$ and $f^\prime(q)\eta(t)$
on the both sides of Eq.(\ref{2.16b}) and rearrange it to obtain

\begin{subequations}
\begin{eqnarray}
\dot{q}&=&p\label{2.23a}\\
\dot{p}&=&- V^\prime(q) +Q_V-\Gamma [f^\prime(q)]^2 p+Q_1+
f^\prime(q) \eta(t)+Q_2\label{2.23b}
\end{eqnarray}
\end{subequations}

Where,

\begin{subequations}
\begin{eqnarray}
Q_V&=&V^\prime(q)-\langle V^\prime(\hat{q}) \rangle\label{2.24a}\\
Q_1&=&\Gamma \left[[f^\prime(q)]^2 p-\langle [f^\prime(\hat{q})]^2
\hat{p}\rangle\right]\label{2.24b}\\
Q_2&=& \eta(t)\left[\langle f^\prime(\hat{q})
\rangle-f^\prime(q)\right]\label{2.24c}
\end{eqnarray}
\end{subequations}

Here $Q_V$ represents quantum correction due to nonlinearity of
the system potential. $Q_1$ and $Q_2$ represent quantum
corrections due to nonlinearity of the system-bath coupling
function. This implies that the quantum Langevin equation is
governed by a c-number noise $\eta(t)$ originating from the heat
bath characterized by the properties (\ref{2.20}-\ref{2.21}) and
the quantum correction terms $Q_V$, $Q_1$ and $Q_2$ are
characteristic of the nonlinearity of the potential and the
coupling function.

Referring to the quantum nature of the system in the Heisenberg
picture we now write the system operators $\hat{q}$ and $\hat{p}$
as

\begin{subequations}
\begin{eqnarray}
\hat{q} & = & q + \delta \hat{q}\label{2.25a}\\
\hat{p} & = & p+\delta \hat{p}\label{2.25b}
\end{eqnarray}
\end{subequations}

where $q(=\langle \hat{q}\rangle)$ and $p(=\langle
\hat{p}\rangle)$ are the quantum mechanical mean values and
$\delta\hat{q}$ and $\delta\hat{p}$ are the operators and they
are quantum fluctuations around their respective mean values. By
construction $\langle \delta\hat{q}\rangle=\langle
\delta\hat{p}\rangle=0$ and they also follow the usual
commutation relation $[\delta\hat{q}, \delta\hat{p}]=i\hbar$.
Using (\ref{2.25a}) and (\ref{2.25b}) in $V^\prime(\hat{q})$,
$[f^\prime(\hat{q})]^2\hat{p}$ and $f^\prime(\hat{q})$ and a
Taylor series expansion in $\delta\hat{q}$ around $q$ it is
possible to express $Q_V$, $Q_1$ and $Q_2$ respectively as
\cite{sm,akp}

\begin{eqnarray}
Q_V & = & -\sum_{n\geq 2} \frac{1}{n!} V^{n+1}(q) \langle \delta
\hat{q}^n \rangle\label{2.26}\\
Q_1 & = & -\Gamma\; [ 2 \;p\; f^\prime
(q)Q_f + p\; Q_3 + 2 f^\prime (q) Q_4 +Q_5]\label{2.27}\\
Q_2 & = &\eta(t)\; Q_f \label{2.28}
\end{eqnarray}

where,

\begin{eqnarray}
Q_f &=& \sum_{n\geq 2}\frac{1}{n!} f^{n+1}(q) \langle \delta
\hat{q}^n \rangle\label{2.29}\\
Q_3 & = & \sum_{m\geq 1}\sum_{n\geq
1}\frac{1}{m!}\frac{1}{n!}f^{m+1}(q)f^{n+1}(q) \langle \delta
\hat{q}^m \delta \hat{q}^n \rangle\label{2.30}\\
Q_4 &=&\sum_{n\geq 1}\frac{1}{n!} f^{n+1}(q) \langle \delta
\hat{q}^n\delta\hat{p} \rangle\label{2.31}\\
Q_5 & = & \sum_{m\geq1}\sum_{n\geq
1}\frac{1}{m!}\frac{1}{n!}f^{m+1}(q)f^{n+1}(q) \langle \delta
\hat{q}^m \delta \hat{q}^n \delta \hat{p}\rangle\label{2.32}
\end{eqnarray}

Using (\ref{2.27}) and (\ref{2.28}) in Eq.(\ref{2.23b}) we have
the c-number quantum Langevin equation in the Markovian limit

\begin{subequations}
\begin{equation}\label{2.33a}
\dot{q}= p
\end{equation}

\begin{eqnarray}
\dot{p}= -V^\prime(q)&+&Q_V-\Gamma [f^\prime(q)]^2p- 2 \Gamma\;
p\;
f^\prime(q) Q_f -\Gamma \;p\; Q_3 - 2 \Gamma f^\prime(q) Q_4 -\Gamma \;Q_5\nonumber\\
&+& f^\prime(q)\;\eta(t) +Q_f \;\eta(t)\label{2.33b}
\end{eqnarray}
\end{subequations}

The quantum Langevin equation is characterized by a classical
force term, $V^\prime$, as well as its correction $Q_V$. The
terms containing $\Gamma$ are nonlinear dissipative terms where
$Q_f$, $Q_3$, $Q_4$ and $Q_5$ are due to associated quantum
contribution in addition to classical nonlinear dissipative term
$\Gamma [f^\prime(q)]^2 p$. The last term in the above equation
refers to a quantum multiplicative noise term in addition to the
usual classical contribution $f^\prime(q) \eta(t)$. It is
therefore easy to recognize the classical limit of the above
equation derived earlier by Lindenberg and Seshadri \cite{lin}.
Furthermore quantum dispersions due to potential and coupling
terms in the Hamiltonian are entangled with nonlinearity. The
quantum noise due heat bath on the other hand is expressed in
terms of the fluctuation-dissipation relation.

\subsection{Quantum correction equations}

The quantum correction terms due to nonlinearity of the potential
and the coupling function in Eq.(\ref{2.33b}) are taken care of
to all orders, in principle. In order to calculate them and the
associated equations explicitly we write the system operators
$\hat{q}$, $\hat{p}$ and the noise operator $\hat{\eta}$ in the
Heisenberg picture as

\begin{subequations}
\begin{eqnarray}
\hat{q} & = & q + \delta \hat{q}\;\;\;\;,\label{2.34a}\\
\hat{p} & = & p+\delta \hat{p}\label{2.34b}\;\;\;\;,\\
\hat{\eta} & = & \eta+\delta \hat{\eta}\;\;\;\;,\label{2.34c}
\end{eqnarray}
\end{subequations}

where $\delta\hat{\eta}$ is the fluctuation in noise around the
mean value $\eta$ and also $\langle\delta\hat{\eta}\rangle=0$. We
then use the above set of equations in operator Langevin equation
in the Markovian limit (Eq.(\ref{2.13a}) and Eq.(\ref{2.13b}))
and subtract the c-number quantum Langevin equation
(Eq.(\ref{2.33a}) and Eq.(\ref{2.33b})) from the resultant to
obtain

\begin{eqnarray}
\dot{\delta\hat{q}}&=&\delta\hat{p}\label{2.35}\\
\dot{\delta\hat{p}}&=&-V^{\prime\prime}(q)\delta\hat{q}-\sum_{n\ge
2}\frac{1}{n!}V^{n+1}(q)\left[\delta\hat{q}^n-\langle\delta\hat{q}^n\rangle\right]\nonumber\\
&-&\Gamma\;\left[2f^\prime(q)f^{\prime\prime}(q)\delta\hat{q}
+2f^\prime(q)\sum_{n\ge2}\frac{1}{n!}f^{n+1}(q)\left[\delta\hat{q}^n-\langle
\delta\hat{q}^n\rangle\right]\right.\nonumber\\
&+&\left.\sum_{m\ge 1}\sum_{n\ge
1}\frac{1}{m!}\frac{1}{n!}f^{m+1}(q)f^{n+1}(q)\left[\delta\hat{q}^m\delta\hat{q}^n-\langle
\delta\hat{q}^m\delta\hat{q}^n\rangle\right]\right]p\nonumber\\
&-&\Gamma\;\left[[f^\prime(q)]^2\delta\hat{p}+2f^\prime(q)\sum_{n\ge1}\frac{1}{n!}f^{n+1}(q)
\left[\delta\hat{q}^n\delta\hat{p}-\langle\delta\hat{q}^n\delta\hat{p}\rangle\right]
\right.\nonumber\\
&+&\left.\sum_{m\ge1}\sum_{n\ge1}\frac{1}{m!}\frac{1}{n!}f^{m+1}(q)f^{n+1}(q)
\left[\delta\hat{q}^m\delta\hat{q}^n\delta\hat{p}-\langle\delta\hat{q}^m\delta\hat{q}^n
\delta\hat{p}\rangle\right]\right]\nonumber\\
&+&\eta(t)\left[f^{\prime\prime}(q)\delta\hat{q}+\sum_{n\ge 2
}\frac{1}{n!}f^{n+1}(q)[\delta\hat{q}^n-\langle\delta\hat{q}^n\rangle]\right]\nonumber\\
&+&\delta\hat{\eta}\left[f^\prime(q)+\sum_{n\ge
1}\frac{1}{n!}f^{n+1}(q)\delta\hat{q}^n\right]\label{2.36}
\end{eqnarray}

We now carry out quantum mechanical average of Eq.(\ref{2.36})
over initial product separable bath states
$\Pi_{j=1}^{\infty}\{|\alpha_j(0)\rangle\}$ to get rid of
$\delta\hat{\eta}$ term. With this the correction equations result in,

\begin{subequations}
\begin{eqnarray}
\dot{\delta\hat{q}}&=&\delta\hat{p}\label{2.37a}\\
\dot{\delta\hat{p}}&=&-V^{\prime\prime}(q)\delta\hat{q}-\sum_{n\ge
2}\frac{1}{n!}V^{n+1}(q)\left[\delta\hat{q}^n-\langle\delta\hat{q}^n\rangle\right]\nonumber\\
&-&\Gamma\;\left[2f^\prime(q)f^{\prime\prime}(q)\delta\hat{q}
+2f^\prime(q)\sum_{n\ge2}\frac{1}{n!}f^{n+1}(q)\left[\delta\hat{q}^n-\langle
\delta\hat{q}^n\rangle\right]\right.\nonumber\\
&+&\left.\sum_{m\ge 1}\sum_{n\ge
1}\frac{1}{m!}\frac{1}{n!}f^{m+1}(q)f^{n+1}(q)\left[\delta\hat{q}^m\delta\hat{q}^n-\langle
\delta\hat{q}^m\delta\hat{q}^n\rangle\right]\right]p\nonumber\\
&-&\Gamma\;\left[[f^\prime(q)]^2\delta\hat{p}+2f^\prime(q)\sum_{n\ge1}\frac{1}{n!}f^{n+1}(q)
\left[\delta\hat{q}^n\delta\hat{p}-\langle\delta\hat{q}^n\delta\hat{p}\rangle\right]
\right.\nonumber\\
&+&\left.\sum_{m\ge1}\sum_{n\ge1}\frac{1}{m!}\frac{1}{n!}f^{m+1}(q)f^{n+1}(q)
\left[\delta\hat{q}^m\delta\hat{q}^n\delta\hat{p}-\langle\delta\hat{q}^m\delta\hat{q}^n
\delta\hat{p}\rangle\right]\right]\nonumber\\
&+&\eta(t)\left[f^{\prime\prime}(q)\delta\hat{q}+\sum_{n\ge 2
}\frac{1}{n!}f^{n+1}(q)[\delta\hat{q}^n-\langle\delta\hat{q}^n\rangle]\right]\label{2.37b}
\end{eqnarray}
\end{subequations}

The operator equations (\ref{2.37a} and \ref{2.37b}) are the basis
of calculation of quantum correction terms $Q_V$, $Q_f$, $Q_3$,
$Q_4$ and $Q_5$ to an arbitrary order. More specifically, we are
to set up the equations for $\langle\delta\hat{q}^2\rangle$,
$\langle\delta\hat{q}\delta\hat{p}+\delta\hat{p}\delta\hat{q}\rangle$,
$\langle\delta\hat{p}^2\rangle$ in the second order and similarly
for third order and so on. They are, in general, coupled and the
infinite set of hierarchy of equations must be truncated after
the desired order for practical purpose. The procedure is exactly
similar to what had been done earlier in the case of
additive noise \cite{skb,db1,db2,bk1,bk2}. Thus upto third order
we may construct the following set of equations from Eq.(\ref{2.37a}
and \ref{2.37a});

\begin{subequations}
\begin{eqnarray}
\frac{d}{dt}\langle\delta\hat{q}^2\rangle&=&\langle\delta\hat{q}\delta\hat{p}+
\delta\hat{p}\delta\hat{q}\rangle\label{2.38a}\\
\frac{d}{dt}\langle\delta\hat{q}\delta\hat{p}+\delta\hat{p}\delta\hat{q}\rangle
&=&-2\chi(q,p)\langle\delta\hat{q}^2\rangle + 2
\langle\delta\hat{q}^2\rangle-\Gamma[f^\prime(q)]^2\langle\delta\hat{q}\delta\hat{p}+
\delta\hat{p}\delta\hat{q}\rangle\nonumber\\
&-&\zeta(q,p)\langle\delta\hat{q}^3\rangle -2\Gamma
f^\prime(q)f^{\prime\prime}(q)\langle\delta\hat{q}^2\delta\hat{p}+
\delta\hat{p}\delta\hat{q}^2\rangle\label{2.38b}\\
\frac{d}{dt}\langle\delta\hat{p}^2\rangle&=& - 2 \Gamma
[f^\prime(q)]^2\langle\delta\hat{p}^2\rangle-\chi(q,p)\langle\delta\hat{q}\delta\hat{p}+
\delta\hat{p}\delta\hat{q}\rangle\nonumber\\
&-&\frac{1}{2}\zeta(q,p)\langle\delta\hat{q}^2\delta\hat{p}+
\delta\hat{p}\delta\hat{q}^2\rangle-2\Gamma
f^\prime(q)f^{\prime\prime}(q)
\langle\delta\hat{q}\delta\hat{p}^2+
\delta\hat{p}^2\delta\hat{q}\rangle\label{2.38c}\\
\frac{d}{dt}\langle\delta\hat{q}^3\rangle&=&\frac{3}{2}\langle\delta\hat{q}^2\delta\hat{p}+
\delta\hat{p}\delta\hat{q}^2\rangle\label{2.38d}\\
\frac{d}{dt}\langle\delta\hat{p}^3\rangle&=&-3 \Gamma
[f^\prime(q)]^2\langle\delta\hat{p}^3\rangle-\frac{3}{2}\chi(q,p)\langle\delta\hat{q}
\delta\hat{p}^2+\delta\hat{p}^2\delta\hat{q}\rangle\label{2.38e}\\
\frac{d}{dt}\langle\delta\hat{q}^2\delta\hat{p}+
\delta\hat{p}\delta\hat{q}^2\rangle &=& -2
\chi(q,p)\langle\delta\hat{q}^3\rangle+2\langle\delta\hat{q}\delta\hat{p}^2+
\delta\hat{p}^2\delta\hat{q}\rangle\nonumber\\
&-&\Gamma[f^\prime(q)]^2\langle\delta\hat{q}^2\delta\hat{p}+
\delta\hat{p}\delta\hat{q}^2\rangle\label{2.38f}\\
\frac{d}{dt}\langle\delta\hat{q}\delta\hat{p}^2+
\delta\hat{p}^2\delta\hat{q}\rangle &=& 2
\langle\delta\hat{p}^3\rangle-4\chi(q,p)\langle\delta\hat{q}^2\delta\hat{p}+
\delta\hat{p}\delta\hat{q}^2\rangle\nonumber\\
&-&2 \Gamma[f^\prime(q)]^2\langle\delta\hat{q}\delta\hat{p}^2
+\delta\hat{p}^2\delta\hat{q}\rangle\label{2.38g}\\
where\nonumber\\
\chi(q,p)&=&V^{\prime\prime}(q)+2\Gamma\;p\;f^\prime(q)
f^{\prime\prime}(q)-\eta(t)f^{\prime\prime}(q)\label{2.38h}\\
\zeta(q,p)&=&V^{\prime\prime\prime}(q)+2\Gamma\;p\;f^\prime(q)
f^{\prime\prime\prime}(q)+2\Gamma\;p\;[f^{\prime\prime}(q)]^2-
\eta(t)f^{\prime\prime\prime}(q)\label{2.38i}
\end{eqnarray}
\end{subequations}

\subsection{Calculation of quantum statistical averages}

Summarizing the discussions of the last two sections $A$ and $B$
we now see that the quantum Langevin dynamics can be calculated
for a stochastic process by solving the coupled Eqs.(2.33a,
2.33b) and (2.38a-2.38g) for quantum mechanical mean values
simultaneously with quantum correction equations which describe
quantum dispersion around these mean values. In principle, the
equations for quantum corrections constitute an infinite set of
hierarchy which must be truncated after a desired order, in
practice, to make the system of equations closed. Care must be
taken to distinguish three averages, the quantum mechanical mean
$\langle\hat{O}\rangle(=O)$, statistical average over quantum
mechanical mean $\langle O \rangle_S$ and the usual quantum
statistical average $\langle\hat{O}\rangle_{QS}$ as discussed in
Sec.A. To illustrate the relation among them let us
calculate, for example, the quantum statistical averages
$\langle\hat{q}\rangle_{QS}$, $\langle\hat{q}^2\rangle_{QS}$ and
$\langle\hat{q}^2\hat{p}\rangle_{QS}$. By (\ref{2.34a}) and
(\ref{2.34b}) we write

\begin{eqnarray}
\hat{q}&=&q+\delta\hat{q}\nonumber\\
\langle\hat{q}\rangle_{QS}&=&\langle {q + \delta\hat{q}}
\rangle_{QS}\label{2.39}\\
&=& \langle q \rangle_{S}+\langle \langle \delta \hat{q} \rangle
\rangle_S=\langle q \rangle_{S}\label{2.40}
\end{eqnarray}

Again

\begin{eqnarray}
\langle\hat{q}^2\rangle_{QS}&=&\langle(q+\delta\hat{q})^2\rangle_{QS}\nonumber\\
&=&\langle q^2 \rangle_S+\langle \langle \delta \hat{q}^2 \rangle
\rangle_S\label{2.41}
\end{eqnarray}

In the case of harmonic potential and linear coupling,
$\langle\delta \hat{q}^2\rangle$ is independent of $q$ or $p$ so
that one may simplify (\ref{2.41}) further as

\begin{equation}
\langle\hat{q}^2\rangle_{QS}=\langle q^2 \rangle_S+\langle \delta
\hat{q}^2 \rangle\label{2.42}
\end{equation}

In Ref. \cite{db2} the explicit exact expressions for $\langle
\hat{q}^2\rangle_{QS}$ and $\langle\hat{p}^2\rangle_{QS}$ have
been derived for harmonic oscillator in linear system-bath
coupling and they are found to be in exact agreement with those
of Grabert \textit{et al} \cite{gra}. For anharmonic potential,
however, one must have to use (\ref{2.41}) to carry out further
the statistical average over $\langle\delta\hat{q}^2\rangle$
\textit{i.e} $\langle\langle\delta\hat{q}^2\rangle\rangle_S$,
since $\langle \delta \hat{q}^2\rangle$ is a function of stochastic
variables $q$ and $p$ according to quantum correction equations.
Furthermore we consider $\langle\hat{q}^2\hat{p}\rangle_{QS}$

\begin{eqnarray}
\langle\hat{q}^2\hat{p}\rangle_{QS}&=&\langle
(q+\delta\hat{q})^2(p+\delta\hat{p}) \rangle_{QS}\nonumber\\
&=&\langle q^2p\rangle_S+\langle p \;\langle \delta\hat{q}^2
\rangle\rangle_S+\langle \langle \delta\hat{q}^2 \;\delta\hat{p}
\rangle\rangle_S+2 \;\langle\; q \;\langle
\delta\hat{q}\;\delta\hat{p}\rangle\rangle_S\label{2.43}
\end{eqnarray}

The essential element of the present approach is thus expressing
the quantum statistical average as the sum of statistical
averages of set of functions of quantum mechanical mean values
and dispersions. Langevin dynamics being coupled to quantum
correction equations, the quantum mechanical mean values as well
as the dispersions are computed simultaneously for each
realization of the stochastic "path". A statistical average implies
the averaging over many such "paths" (typically three to five
thousands) similar to what is done to calculate statistical
averaging by solving classical Langevin equation. Before leaving
this section we mention a few pertinent points.

First, the distinction between the ensemble averaging by the
present procedure and by the standard approach using Wigner
function is now clear. From Eq.(\ref{2.43}) we note that, for
example,

\begin{equation}\label{2.44}
\langle \hat{q}^2 \hat{p} \rangle_{QS}=\int q^2 p\; W(q,p)\neq
\langle q^2 p \rangle_S
\end{equation}

where $W(q,p)$ is the Wigner function for the system. (This is
not be confused with the Wigner function we introduced in
Eq.(\ref{2.18}) for the bath oscillators).

Second, our formulation of the Langevin equation coupled to
quantum correction equations belongs to quantum stochastic process
driven by c-number  noise, which is classical-like in form. Its
numerical solutions can be obtained \cite{db2,bk1} in the same
way as one proceeds in a classical theory.

Third, quantum nature of the dynamics appears in two different
ways. The heat bath is quantum mechanical in character whose noise
properties are expressed through quantum fluctuation-dissipation
relation. The nonlinearity of the system potential and coupling,
on the other hand, give rise to quantum correction terms. Thus
the classical Langevin equation can be easily recovered (i) in the
limit $\hbar\omega\ll k_BT$ to be applied in the Eq.(\ref{2.21})
so that one obtains the classical fluctuation-dissipation relation
and (ii) if the quantum dispersion terms vanish.

\section{The overdamped limit and the stationary distribution}

\subsection{The Langevin equation with multiplicative noise
under strong friction}

In the case of large dissipation, one eliminates the fast
variables adiabatically to get a simpler description of the
system which is valid in a much slower time scale. This adiabatic
elimination of fast variables is basically a zeroth order
approximation. In this zeroth order approximation the number of
system variables get reduced. When the Brownian particles move in
a bath with constant large dissipation this adiabatic elimination
of fast variables leads to the correct description of the system.
This wellknown approximation, known as Smoluchowski
approximation, results in correct equilibrium distribution.
However in presence of hydrodynamic interaction, \textit{i.e.},
when the fluctuation is position/state dependent or equivalently
when the noise is multiplicative with respect to system variables
the conventional adiabatic reduction of fast variables does not
give the correct description of the system. Several years ago
Sancho \textit{et al} \cite{san1} had proposed an alternative
approach to get a correct Langevin equation in the case
multiplicative noise system. Based on the Langevin equation they
carried out a systematic expansion of the relevant variables in
powers of $\Gamma^{-1}$ neglecting terms smaller than
$O(\Gamma^{-1})$. Then by ordinary Stratonovich interpretation it
is possible to obtain the correct Langevin equation corresponding
to a Fokker-Planck equation in position space. This description
leads to the correct stationary probability distribution of the
system with position dependent friction.

In order to get the quantum Langevin equation in the overdamped limit we
follow this procedure. In what follows we discard the quantum
correction terms $Q_4$ and $Q_5$ from Eq.(\ref{2.33b}) since
they involve quantum dispersions $\delta\hat{p}$ and therefore
decays exponentially in the large damping limit. In this limit
these transient correction terms do not affect the dynamics of
the position which varies in a much slower time scale. So the
quantum Langevin equation Eq.(\ref{2.33a}) and Eq.(\ref{2.33b}) can be
written as, respectively,

\begin{eqnarray}
\dot{q}&=&p\label{3.1}\\
\dot{p}&=&-V^\prime(q)+Q_V-\Gamma\; h(q) p+g(q)\;
\eta(t)\label{3.2}
\end{eqnarray}

where

\begin{eqnarray}
h(q) &=& [f^\prime(q)]^2+2 f^\prime(q) Q_f +Q_3\label{3.3}\\
g(q) &=& f^\prime(q)+Q_f\label{3.4}
\end{eqnarray}

The method of Sancho \textit{et al} \cite{san1} is followed
further to obtain the Fokker-Planck equation in position space
corresponding Langevin equation Eq.(\ref{3.2})

\begin{eqnarray}
\frac{\partial P(q,t)}{\partial t}&=&\frac{\partial}{\partial
q}\left[ \frac{V^\prime(q)-Q_V}{\Gamma\; h(q)} \right]
P(q,t)+D_0\frac{\partial}{\partial
q}\left[\frac{1}{\Gamma\;(h(q))^2}\;
g^\prime(q) g(q) \right] P(q,t)\nonumber\\
&+& D_0\frac{\partial}{\partial
q}\left[\frac{g(q)}{\Gamma\;h(q)}\;\frac{\partial}{\partial
q}\;\frac{g(q)}{\Gamma\;h(q)}\right]P(q,t)\label{3.5}
\end{eqnarray}

In the ordinary Stratonovich description the Langevin equation
corresponding to the Fokker-Planck Eq.(\ref{3.5}) is given by

\begin{equation}\label{3.6}
\dot{q}=-\;\frac{V^\prime(q)-Q_V}{\Gamma\;h(q)}-D_0\;\frac{g^\prime(q)
g(q)}{\Gamma\;(h(q))^2}\; +\frac{g(q)}{\Gamma\;h(q)}\;\eta(t)
\end{equation}

Eq.(\ref{3.6}) is c-number quantum Langevin equation for
multiplicative noise with position dependent friction in the
overdamped limit (\textit{i.e.}, corrected upto $O(1/\Gamma)$).

In the classical limit, \textit{i.e.}, $\hbar\omega_0 \ll k_B T$,
$h(q)=[f^\prime(q)]^2$, $g(q)=f^\prime(q)$, $Q_V=0$ and
$D_0=\Gamma\;k_B T$ so the c-number quantum Langevin equation
reduces to
\begin{equation}\label{3.7}
\dot{q}=\frac{1}{\Gamma\;[f^\prime(q)]^2}
\left[-V^\prime(q)-\Gamma\;k_B T\;
\frac{f^{\prime\prime}(q)}{f^\prime(q)}+f^\prime(q)\;\eta(t)
\right]
\end{equation}

which is exactly the form derived by Sancho \textit{et al}
\cite{san1}.

Our next task is to establish the quantum corrections in the
overdamped limit. To this end we return to Eq.(\ref{2.37b}),
neglect the $\delta\dot{\hat{p}}$ term and keep the leading order
correction equation (since in the overdamped limit higher order
quantum contributions are small) to obtain

\begin{equation}\label{3.8}
\frac{d}{dt}\;\delta\hat{q}=\frac{1}{\Gamma\;[f^\prime(q)]^2}\left[-V^{\prime\prime}(q)\;
\delta\hat{q}-2\Gamma\;p\;f^\prime(q)f^{\prime\prime}(q)\delta\hat{q}+\eta(t)
f^{\prime\prime}(q)\delta\hat{q}\right]+O(\delta\hat{q}^2)
\end{equation}

From Eq.(\ref{3.8}) it is easy to calculate the equations of
motion for quantum correction in the lowest order as

\begin{eqnarray}\label{3.9}
\frac{d}{dt}\langle\delta\hat{q}^2\rangle=\frac{2}{\Gamma\;[f^\prime(q)]^2}\left[
-V^{\prime\prime}(q)\langle\delta\hat{q}^2\rangle-2\;\Gamma\;p\;f^\prime(q)
f^{\prime\prime}(q)\langle\delta\hat{q}^2\rangle+\eta(t)
f^{\prime\prime}(q)\langle\delta\hat{q}^2\rangle\right]
\end{eqnarray}

A simplified expression for the leading order quantum correction
term $\langle\delta\hat{q}^2\rangle$ can be estimated by
neglecting the higher order coupling terms in the square bracket
in Eq.(\ref{3.9}) and rewriting it as $d\langle\delta\hat{q}^2
\rangle=\frac{2}{\Gamma\;[f^\prime(q)]^2}\;V^{\prime\prime}(q)
\langle\delta\hat{q}^2\rangle \; dt$. The overdamped
deterministic classical motion on the other hand gives
$dq=-\;\frac{V^\prime(q)}{\Gamma\;[f^\prime(q)]^2}\;dt$. These
together yield after integration

\begin{equation}\label{3.10}
\langle\delta\hat{q}^2\rangle=\Delta_q [V^\prime(q)]^2
\end{equation}

where
$\Delta_q=\frac{\langle\delta\hat{q}^2\rangle_0}{[V^\prime(q_0)]^2}$
and $q_0$ refers to initial position.

\subsection{The stationary solution of Fokker-Planck equation in overdamped limit}

The Fokker-Planck equation Eq.(\ref{3.5}) in the overdamped limit
can be rewritten in a more compact form as

\begin{equation}\label{3.11}
\frac{\partial P(q,t)}{\partial t}=\frac{\partial}{\partial
q}\;\frac{1}{\Gamma\;h(q)}\left[V^\prime(q)-Q_V+\frac{D_0}{\Gamma}\;\frac{\partial}{\partial
q }\;\frac{g(q)^2}{h(q)}\right]P(q,t)
\end{equation}

Eq.(\ref{3.11}) is the quantum Smoluchowski equation for
multiplicative noise (damping is corrected upto $O(1/\Gamma)$). To
capture the essential content of state-dependent diffusion we now
proceed as follows. Under the stationary condition

\begin{equation}\label{3.12}
\frac{\partial P(q,t)}{\partial t}==0
\end{equation}

Eq.(\ref{3.11}) reduces to

\begin{equation}\label{3.13}
\frac{D_0}{\Gamma}\;\frac{d}{d q
}\left[\frac{g(q)^2}{h(q)}\;P_{st}(q)\right]+V^\prime(q)-Q_V=0
\end{equation}

After integration of Eq.(\ref{3.13}) we have the stationary
probability distribution in the overdamped limit as

\begin{equation}\label{3.14}
P_{st}(q)=N\frac{1}{[g(q)^2/h(q)]}\;\exp\left[-\int_0^q\frac{V_{quan}^\prime(q^\prime)}
{D(q^\prime)}\;dq^\prime\right]
\end{equation}

with

\begin{eqnarray*}
D(q)&=&\hbar\omega_0
\left(\bar{n}(\omega_0)+\frac{1}{2}\right)\left[g(q)^2/h(q)\right]\\
\end{eqnarray*}

and

\begin{eqnarray*}
V_{quan}^\prime &=& V^\prime(q)-Q_V
\end{eqnarray*}

and $N$ is the normalization constant. In the classical limit we
have $g(q)^2/h(q)=1$, $Q_V=0$ and $\hbar\omega_0
(\bar{n}(\omega_0)+\frac{1}{2})=k_B T$; the stationary
probability distribution function (\ref{3.14}) reduces to
classical equilibrium Boltzmann distribution
$P^c_{st}=N\exp[-V(q)/k_B T]$. Therefore the stationary
distribution (\ref{3.14}) is essentially a quantum generalization
of Boltzmann factor for state-dependent diffusion. This diffusion
arises where the inhomogeneity is due to quantum corrections
entangled with nonlinearity of the system-bath coupling. The
state-dependent diffusion is wellknown in several classical
contexts as pointed out earlier notably in Landauer Blow-torch
effect \cite{lan1,lan2}, where localized heating in a region
along the reaction coordinate lying between the lower energy
minimum and barrier top can raise the relative population of the
higher minimum above what is allowed by Boltzmann factor. The
example also includes noise-induced transport processes due to
phase difference between periodic modulation of drift and
modulation of diffusion as suggested by B\"{u}ttiker \cite{but}.
It is however worth-noting that the inhomogeneity of $D(q)$ as
implied in Eq.(\ref{3.14}) is reminiscent of some sort of quantum
nonlocal effect. A number of points are pertinent here. First, in
realizing inhomogeneity through $D(q)$ we have taken care of
quantum corrections to all orders. Second, the state-dependence
is essentially due to nonlinear coupling mechanism in Hamiltonian
(\ref{2.1}) and is therefore model dependent and because of
c-number nature of our treatment it corresponds to one of the
classical forms of the phenomenological state-dependent diffusion
term. This correspondence makes classical-quantum correspondence
more clear. It is also worthnoting that the model-dependent
nature of escape rate in classical Blow-torch effect had been
discussed recently \cite{sol}. Third, for harmonic potential and
linear coupling $f(q)=q$, we have $Q_V=0$ and $g(q)^2/h(q)=1$ and
the probability distribution function reduces to well-known
Wigner distribution for harmonic oscillator $V(q)$ as
$P^h_{st}(q)=N\exp[-V(q)/\hbar\omega_0
(\bar{n}(\omega_0)+\frac{1}{2})]$. Fourth, although
model-dependent the inhomogeneous nature of the quantum diffusion
must be thermodynamically consistent. To this end we examine in
the next section the equilibrium condition as well as typical
nonzero current situation (due to symmetry breaking under special
condition) for a periodic potential and periodic derivative of
coupling function. By thermodynamic consistency we imply that in
absence of any external field, no directional component should
remain after appropriate averaging over ensemble or over the
period of space or time.

\section{Application: A Periodic potential}

\subsection{Solution under periodic boundary condition; Thermodynamic consistency}

In the overdamped limit the stationary current from Eq.(3.11) can
be represented as,

\begin{equation}\label{4.1}
J=-\;\frac{1}{\Gamma\;h(q)}\left[V^\prime(q)-Q_V+\frac{D_0}{\Gamma}\;\frac{d}{dq}
\left(\frac{g(q)^2}{h(q)}\right)\right]P_{st}(q)
\end{equation}

Integrating the Eq.(\ref{4.1}) we have the expression of
stationary probability distribution in terms of stationary
current as,

\begin{equation}\label{4.2}
P_{st}(q)=\frac{e^{-\phi(q)}}{[g(q)^2/h(q)]}\left[\frac{g(0)^2}{h(0)}\;
P_{st}(0)-J\;\frac{\Gamma^2}{D_0}\int_0^q
h(q^\prime)e^{\phi(q^\prime)}dq^\prime\right]
\end{equation}

where

\begin{equation}\label{4.3}
\phi(q)=\frac{\Gamma}{D_0}\int_0^q\frac{V^\prime(q^\prime)-Q_V}{[g(q^\prime)^2/h(q^\prime)]}\;
dq^\prime
\end{equation}

We now consider a symmetric periodic potential with periodicity
$2\pi$, \textit{i.e.}, $V(q)=V(q+2\pi)$ and periodic derivative
of coupling function with the same periodicity as that of the
potential, \textit{i.e.}, $f^\prime(q)=f^\prime(q+2\pi)$.

Since the potential is periodic, $Q_V$ is also a periodic
function because $Q_V=V(q)-\langle V(\hat{q})\rangle$. Similarly
$Q_f$ is also periodic because $Q_f=\langle
f^\prime(\hat{q})\rangle -f^\prime(q)$, and also
$Q_3+2f^\prime(q)Q_f$ is also periodic since
$Q_3+2f^\prime(q)Q_f=\langle
[f^\prime(\hat{q})]^2\rangle-[f^\prime(q)]^2$. So from
Eqs.(\ref{3.3}) and (\ref{3.4}) it is clear that $h(q)$ and
$g(q)$ are also periodic functions of $q$ with periodicity $2\pi$.
This will be made more explicit when we consider a specific
example in the next section.

Now applying the periodic boundary condition on $P_{st}(q)$,
\textit{i.e.}, $P_{st}(q)=P_{st}(q+2\pi)$ we have from
Eq.(\ref{4.2})

\begin{equation}\label{4.4}
\frac{g(0)^2}{h(0)}\;
P_{st}(0)=J\;\frac{\Gamma^2}{D_0}\left[1-e^{\phi(2\pi)}\right]^{-1}\int_0^{2\pi}
h(q)\;e^{\phi(q)}dq
\end{equation}

By applying the normalization condition on stationary probability
distribution which is given by,

\begin{equation}\label{4.5}
\int_0^{2\pi}P_{st}(q)\;dq=1
\end{equation}

we obtain from Eq.(\ref{4.2})

\begin{equation}\label{4.6}
\int_0^{2\pi}\frac{e^{-\phi(q)}}{[g(q)^2/h(q)]}\left[\frac{g(0)^2}{h(0)}\;
P_{st}(0)-J\;\frac{\Gamma^2}{D_0}\int_0^q
h(q^\prime)\;e^{\phi(q^\prime)}dq^\prime\right]\;dq=1
\end{equation}

Elimination of $\frac{g(0)^2}{h(0)}\; P_{st}(0)$ from (\ref{4.4})
and (\ref{4.6}) yields the expression of stationary current after
some rearrangement

\begin{equation}\label{4.7}
J=\frac{D_0}{\Gamma^2}\;\frac{1-e^{\phi(2\pi)}}{\int_0^{2\pi}\frac{h(q)}{g(q)^2}
\;e^{-\phi(q)}dq\int_0^{2\pi}h(q^\prime)\;e^{\phi(q^\prime)}dq^\prime
-[1-e^{\phi(2\pi)}]\int_0^{2\pi}\frac{h(q)}{g(q)^2}\;e^{-\phi(q)}\int_0^q
h(q^\prime)\;e^{\phi(q^\prime)}dq^\prime dq}
\end{equation}

From the condition of periodicity of potential and different
quantum correction terms it is clear that for the periodic
potential and the periodic derivative of coupling function with
same periodicity $\frac{V^\prime(q)}{[g(q)^2/h(q)]}$ and
$\frac{Q_V(q)}{[g(q)^2/h(q)]}$ are both periodic with same
periodicity. This makes effective potential $\phi(q)$ equal to
zero;

\begin{equation}\label{4.8}
\phi(2\pi)=\frac{\Gamma}{D_0}\int_0^{2\pi}\frac{V^\prime(q)-Q_V}{[g(q)^2/h(q)]}\;dq=0
\end{equation}

Therefore the numerator of the expression for current (\ref{4.7})
reduces to zero. We thus conclude that there is no occurrence of
current for a periodic potential and periodic derivative of coupling
with same periodicity. At the macroscopic level this confirms that there is
no generation of perpetual motion of second kind, \textit{i.e.},
no violation of second law of thermodynamics. Therefore the
thermodynamic consistency based on symmetry considerations
ensures the validity of the present formalism and of the overdamped
quantum multiplicative Langevin equation.

\subsection{Phase induced current}

B\"{u}tikker \cite{but} in 1987 had shown that in case of
space-dependent friction in the overdamped limit a classical
particle under a symmetric sinusoidal potential field and also in
presence of a sinusoidally modulated space-dependent diffusion
with same periodicity experiences a net drift force resulting in
generation of current. This current is basically due to the phase
difference between the symmetric periodic potential and the space
dependent diffusion. The current does vanish when the phase
difference is either zero or integral multiple of $\pi$. van
Kampen \cite{van} in a later work also ended up with similar kind
of conclusion for a system with space dependent temperature under
the overdamped condition. The result of van Kampen is a
re-examination of the earlier observation due to Landauer
\cite{lan1,lan2}.

We now explore, in the spirit of B\"{u}ttiker but in the present
quantum-mechanical context a system with a symmetric periodic
potential, \textit{i.e.}, $V(q)=V(-q)$, and a periodic derivative
of coupling function with same periodicity but with a phase
difference between them leading to a net directed motion or
current. This is because of the fact that the phase bias gives a
tilt to the effective potential $\phi(q)$, (if $\phi(q)$ is
plotted as a function of q), which makes the transition between
left to right and right to left unequal. The phase difference
therefore breaks the detailed balance of the system.
We show that when the phase difference is zero or
integral multiple of $\pi$ the quantum current vanishes.

We proceed with a sinusoidal symmetric potential of the form

\begin{equation}\label{4.9}
V(q)=V_0\left[1+\cos(q+\theta)\right]
\end{equation}

where $V_0$ can be taken as the barrier height and $\theta$ is
phase factor which can be controlled externally.

The derivative of the chosen coupling function, $f(q)=(q+\alpha \;
\sin q) $, is

\begin{equation}\label{4.10}
f^\prime(q)=1+\alpha\cos q
\end{equation}

where $\alpha$ is a modulation parameter.

Given the form of potential $V(q)$ and coupling function $f(q)$
we obtain from Eq.(\ref{3.10}) the second order quantum correction
in the overdamped limit as

\begin{equation}\label{4.11}
\langle \delta \hat{q}^2 \rangle=-\;\Delta_q\; V_0^2
\sin^2(q+\theta)
\end{equation}

Therefore the correction to the potential in the leading order
from Eq.(\ref{2.26}) is given by

\begin{equation}\label{4.12}
Q_V=-\;\frac{1}{2}\;\Delta_q\; V_0^3 \sin^3(q+\theta)
\end{equation}

The quantum corrections $Q_f$ and $Q_3$ in the same order can be
estimated using Eqs.(\ref{2.29}) and (\ref{2.30}):

\begin{eqnarray}
Q_f &=& -\;\frac{1}{2}\;\Delta_q \;\alpha\; V_0^2 \; \cos q\;
\sin^2(q+\theta)\label{4.13}\\
Q_3 &=& \Delta_q\; \alpha^2\; V_0^2 \; \sin^2q\;
\sin^2(q+\theta)\label{4.14}
\end{eqnarray}

Furthermore from Eqs.(\ref{3.3}) and (\ref{3.4}) we calculate
$h(q)$ and $g(q)$, respectively, using Eqs.(\ref{4.10}),
(\ref{4.13}) and (\ref{4.14}) to obtain

\begin{eqnarray}
h(q)=(1+\alpha\;\cos q)^2&-&\Delta_q\; \alpha\; V_0^2\;\cos
q\;\sin^2(q+\theta)\;(1+\alpha\;\cos q)\nonumber\\
&+&\Delta_q\;\alpha^2\; V_0^2 \; \sin^2q\;
\sin^2(q+\theta)\label{4.15}
\end{eqnarray}

\begin{eqnarray}\label{4.16}
g(q)=1+\alpha\;\cos q-\frac{1}{2}\;\Delta_q\; \alpha\; V_0^2 \;
\cos q\; \sin^2(q+\theta)
\end{eqnarray}

With these preliminaries we now calculate the current given by
Eq.(\ref{4.7}) explicitly using (\ref{4.9}), (\ref{4.12}),
(\ref{4.15}) and (\ref{4.16}). A key quantity for this analysis
is $\phi(q)$ [(\ref{4.3})] which involves the phase $\theta$. In
units of $\hbar=k_B=1$ we set the parameter values $\langle
\delta \hat{q}^2 \rangle_0=1/2$, minimum uncertainty value,
$\Delta_q=0.5$, $V_0=1.0$, $\omega_0=1.0$, $\alpha=1.0$, $T=1.0$
and $\Gamma=1.0$. The variation of current $J$ as a function of
phase $\theta$ is exhibited in Fig.1. It is interesting to
observe that for $\theta\neq 0, n \pi$ where $n=\pm 1, \pm 2...$,
phase induces a current which is a periodic function of the phase
difference between modulations of potential and diffusion. The
origin of the current can be traced to the inhomogeneous
diffusion of a quantum particle in contrast to classical one
proposed by B\"{u}ttiker. That this current depends on the
amplitude of modulation $\alpha$ of the diffusion is shown in
Fig.2 for $\theta=3.6$, $V_0=1.0$, $\omega_0=1.0$ for several
values of $k_B T$. For $\alpha=0$ the current vanishes and we
have only linear coupling and the phase bias has no relevance
in such situation.
We observe for a fixed temperature a sharp
increase in current beyond a moderate value of $\alpha$ and that
for an optimum temperature the current is maximum for a given
strength of modulation. Fig.3 illustrates the variation of
current ($J$) as a function of temperature for the phase
$\theta=3.6$ and for several values of strength of modulation of
diffusion $\alpha$ for $V_0=1.0$, $\omega_0=1.0$. One observes
that even at $T=0$, the vacuum field of the heat bath induces a
finite current. As the temperature increases the current
decreases after an initial increase and reaching a maximum.

We thus observe that nonlinear system-bath coupling may give rise
to state-dependent noise and diffusion in a quantum system. For a
periodic potential and for a periodic derivative of coupling function,
with same periodicity, the state-dependent noise may lead to symmetry
breaking in presence of a phase bias. This generates a directional
flow which vanishes in absence of the bias and can be optimized by
application of suitable strength of modulation and temperature.

\section{Conclusion}

In this paper we have developed a theory of diffusion of a
quantum particle in inhomogeneous media. The approach is based on
the system-reservoir model with a nonlinear coupling. We derive
the quantum Langevin equation with a multiplicative noise and a
nonlinear dissipation in the Markovian limit, which is coupled to
a set of quantum correction equations developed order by order. A
systematic expansion of the relevant variable in powers of
inverse of the dissipation constant and use of large friction
limit lead to a quantum Smoluchowski equation for state-dependent
diffusion. It is apparent that the state dependence owes its
origin to nonlinear coupling between the system and bath degrees
of freedom and the corresponding generalization of Boltzmann
factor for the steady state has been shown to be
thermodynamically consistent. We have applied the formalism to
the problem of diffusion of a quantum particle in a periodic
potential where the derivative of coupling function is periodic
with same periodicity. We have shown that a phase difference
between these two spatially periodic modulations may give rise to
a directed quantum current. This current vanishes in the
classical limit and is a consequence of state-dependent diffusion
where nonlocality in the effective potential is essentially a
quantum effect.

The Brownian motion of a quantum particle in inhomogeneous media
is an active area of contemporary research in stochastic
energetics. We hope that our formalism of quantum Langevin
equation with multiplicative noise and the associated
Smoluchowski equation for state-dependent diffusion as a
description of the stochastic processes may be useful for other
similar issues, particularly for calculations of quantum decay
rate of metastable state, energy dissipation and so on. The
extension of the theory to non-Markovian and weak friction regime
is also worth-pursuing. We hope to address these issues elsewhere.

{\bf Acknowledgements:}\\

We thank Prof. M. B\"{u}ttiker for his kind interest in this work
and Dr. S. K. Banik for interesting discussions. Thanks are also
due to Prof. Y. Tanimura for his critical comments and
discussions. The authors are indebted to the Council of
Scientific and Industrial Research for partial financial support
under Grant No. 01/(1740)/02/EMR-II.

\newpage

\begin{center}
{\bf Figure Captions}
\end{center}

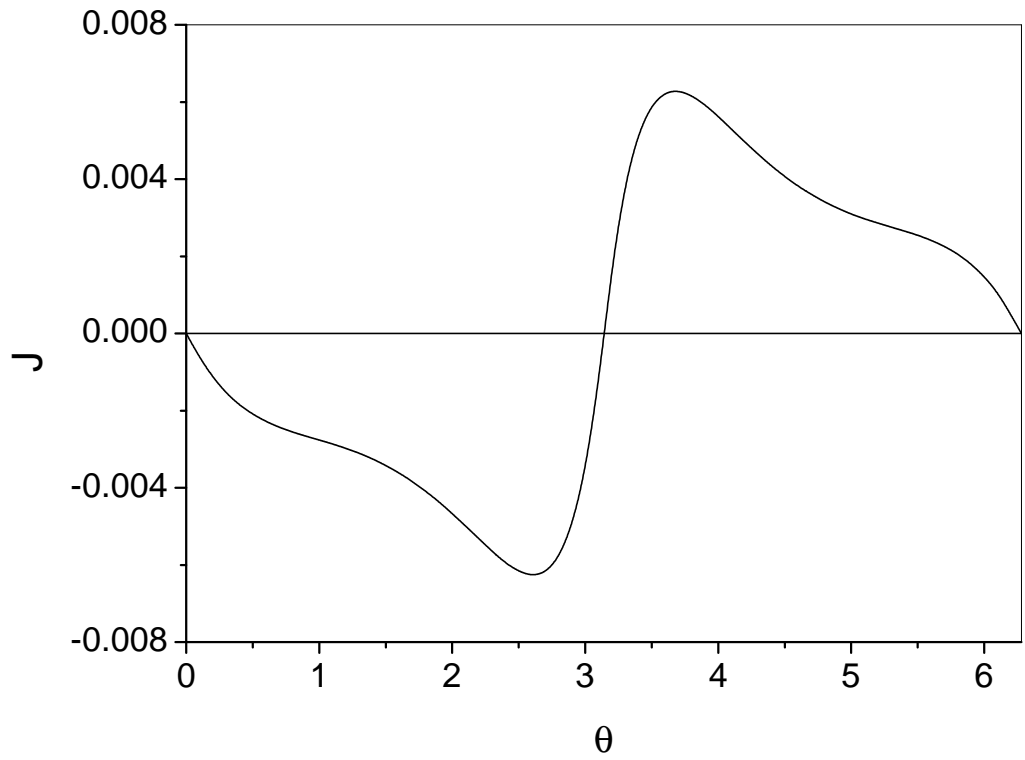
Fig.1: Variation of current $J$ as a function of phase $\theta$
between $0$ to $2\pi$ for $T=1.0$, $\omega_0=1.0$, $\alpha=1.0$,
$V_0=1.0$ and $\Gamma=1.0$ (units arbitrary).

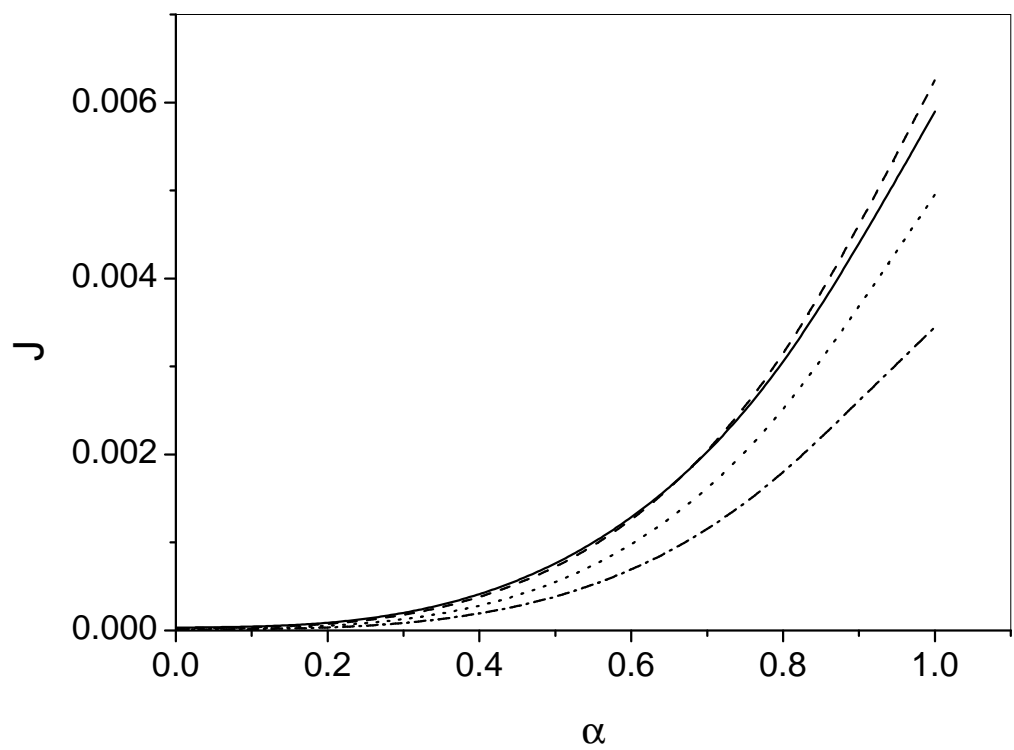
Fig.2: Plot of current $J$ vs. strength of modulation $\alpha$
for $\theta=3.6$, $V_0=1.0$, $\omega_0=1.0$ and for $T=0.1$
(dashed-dotted line); $T=0.5$ (dotted line); $T=1.0$ (dashed
line); $T=3.0$ (solid line) (units arbitrary).

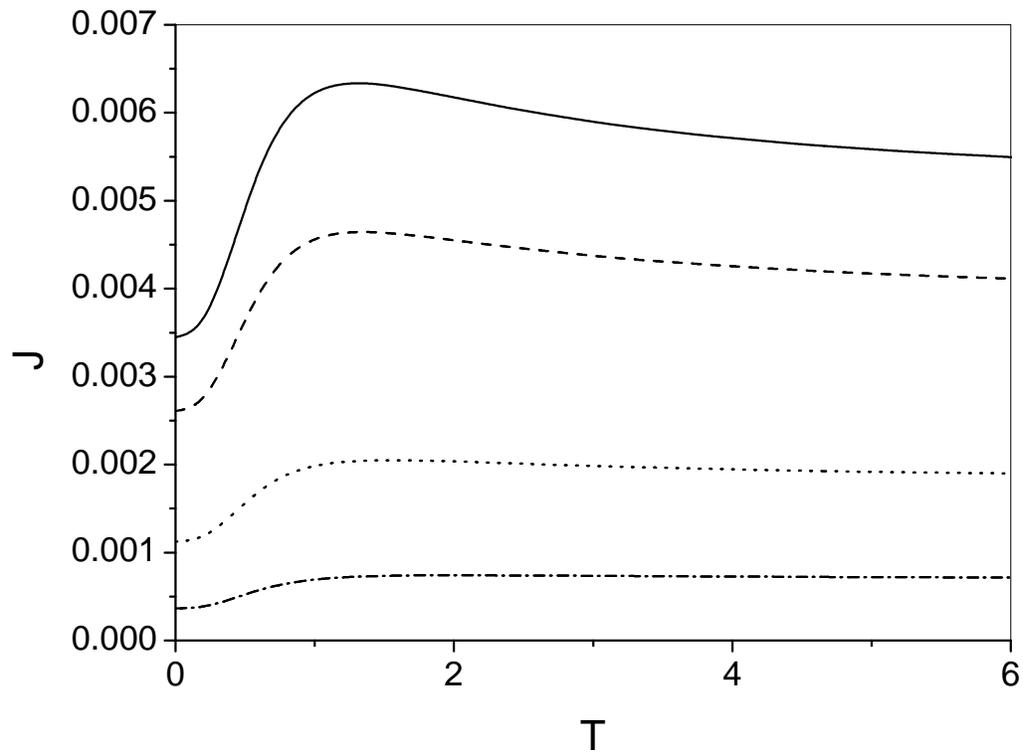
Fig.3: Variation of current $J$ as a function of temperature $T$
for $\theta=3.6$, $V_0=1.0$, $\omega_0=1.0$ and for strength of
modulation $\alpha=0.5$ (dashed-dotted line); $\alpha=0.7$
(dotted line); $\alpha=0.9$ (dashed line); $\alpha=1.0$ (solid
line) (units arbitrary).


\begin{thebibliography}{200}
\bibitem{lan1} R. Landauer, Phys. Rev. A {\bf 12}, 636-638 (1975).

\bibitem{lan2} R. Landauer, J. Stat. Phys. {\bf 53}, 233-248 (1988).

\bibitem{van} N. G. van Kampen, IBM J. Res. Dev. {\bf 32}, 107-111
(1988).

\bibitem{but} M. B\"{u}ttiker, Z. Phys. B: Condensed Matter {\bf
68}, 161-167 (1987).

\bibitem{kuz} P. I. Kuznetsov, R. L. Stratonovich and V. I.
Tikhonov, Sov. Phys. JETP {\bf 1}, 510 (1955); R. L. Stratonovich,
\textit{Topics in the theory of random noise} (Gordon and Breach,
London, 1967).

\bibitem{baut1} Ya. M. Blanter and M. B\"{u}ttiker, Phys. Reps. {\bf
336}, 1-166 (2000).

\bibitem{hak} H. Haken, Rev. Mod. Phys. {\bf 47}, 67-121 (1977).

\bibitem{mar} A. H. Marshak and D. Assaf III, Solid State
Electron. {\bf 16}, 675-679 (1975); P. T. Landsberg and S. A.
Hope, Solid State Electron. {\bf 19}, 173-174 (1977).

\bibitem{mag} M. O. Magnasco, Phys. Rev. Lett. {\bf 71}, 1477-1481 (1993).

\bibitem{jul} F. J\"{u}licher, A. Ajdari and J. Prost, Rev. Mod. Phys.
{\bf 69}, 1269-1282 (1997).

\bibitem{rei} P. Reimann, Phys. Reps. {\bf 361},
57-265 (2002); P. Reimann, M. Grifoni, and P. H\"{a}nggi, Phys.
Rev. Lett. {\bf 79}, 10-13 (1997).

\bibitem{kla} M. Porto, M. Urbakh and J. Klafter,
Phys. Rev. Lett. {\bf 85}, 491-494 (2000); G. Oshanin, J. Klafter,
M. Urbakh and M. Porto, Europhys. Lett. {\bf 68}, 26-32 (2004).

\bibitem{sol} S. Fekade and M. Bekele, Euro. Phys. J. B {\bf 26}, 369-374 (2002).

\bibitem{lin} K. Lindenberg and V. Seshadri, Physica A, {\bf 109}, 483-499 (1981);
K. Lindenberg and E. Cort\'{e}s, \textit{ibid.} {\bf 126}, 489-503
(1984).

\bibitem{pol} E. Pollak and A. M. Berezhkovskii, J. Chem. Phys. {\bf 99},
1344-1346 (1993).

\bibitem{san} J. M. Sancho, M. S. Miguel, S. L. Katz and J. D.
Gunton, Phys. Rev. A {\bf 26}, 1589-1609 (1982).

\bibitem{san1} J. M. Sancho, M. S. Miguel and D. D\"{u}rr, J.
Stat. Phys. {\bf 28}, 291-305 (1982).

\bibitem{jay} A. M. Jayannavar and M. C. Mahato, Pramana {\bf 45},
368-376 (1995).

\bibitem{san2}A. Hernández-Machado, M. S. Miguel and J. M. Sancho,
Phys. Rev. A {\bf 29}, 3388-3396 (1984).

\bibitem{mas} J. Masoliver and L. Garrido, Phys. Lett. A {\bf
103}, 366-368 (1984)

\bibitem{lin1} J. D. Ramshaw and K. Lindenberg, J. Stat. Phys. {\bf
45}, 295-307 (1986).

\bibitem{sak} H. Sakaguchi, J. Phys. Soc. Jpn. {\bf 70}, 3247-3250
(2001).

\bibitem{tai} C. Anteneodo and C. Tsallis, J. Math. Phys. {\bf
44}, 5194-5203 (2003).

\bibitem{san3} J. Garcia-Ojalvo and J. M. Sancho, \textit{Noise in
spatially extended systems} (Springer-Verlag, New York, 1999).

\bibitem{rate} F. Marchesoni, Chem. Phys. Lett. {\bf 110}, 20-24 (1984);
F. de Pasquale, J. M. Sancho, M. S. Miguel and P. Tartaglia ,
Phys. Rev. A {\bf 33}, 4360-4366 (1986); F. Marchesoni, L.
Gammaitoni and E. Menichella-Saetta and S. Santucci, Phys. Lett.
A {\bf 201}, 275-280 (1995); P. Silvestrini, J. Appl. Phys. {\bf
68}, 663-667 (1990); R. Chac\'{o}n, F. Balibrea and M. A.
L\'{o}pez, Phys. Lett. A {\bf 279}, 38-46 (2001); J. B. Straus and
G. A. Voth, J. Chem. Phys. {\bf 96}, 5460-5470 (1992); J. B.
Straus, J. M. Gomez Lorente and G. A. Voth, J. Chem. Phys. {\bf
98}, 4082-4097 (1993).

\bibitem{stka} A. V. Barzykin and K. Seki, Euro. Phys. Lett. {\bf 40},
117-121 (1997); B. Xu, J. Li and J. Jheng, Physica A {\bf 343},
156-166 (2004); L. Gammaitoni, P. H\"{a}nggi, P. Jung and F.
Marchesoni, Rev. Mod. Phys. {\bf 70}, 223-287 (1998).

\bibitem{les} M. R. Young and S. Singh, Opt. Lett. {\bf 13}, 21-23
(1988); G. S. Liu, Opt. Commun. {\bf 79}, 402-406 (1990); X. Zhou,
W. Gao and S. Zhu, Phys. Lett. A {\bf 213}, 43-48 (1996); Q. Long,
L. Cao, D. Wu and Z Li, Phys. Lett. A {\bf 231}, 339-343 (1997).

\bibitem{sig} A. M. Tekalp and G. Pavlovic, IEEE T. Signal Proces.
{\bf 39}, 2132-2136 (1991); L. H. Song and T. S. Uhm, {\bf 138},
531-538 (1991); O. V. Gerashchenko, S. L. Ginzburg, and M. A.
Pustovoit, JETP Lett. {\bf 67}, 997-1003 (1998).

\bibitem{baut} Ya. M. Blanter and M. Büttiker, Phys. Rev. Lett. {\bf 81}
, 4040-4043 (1998).

\bibitem{tran} J. D. Bao, Y. Z. Zhuo and X. Z. Wu, Phys. Lett. A
{\bf 217}, 241-247 (1996); R. Krishnan, M. C. Mahato, and A. M.
Jayannavar, Phys. Rev. E {\bf 70}, 021102-021107 (2004); J. -H.
Li, Y. X. Han and S. G. Chen, Physica D, {\bf 195}, 67-76 (2004).

\bibitem{tran1} W. Horsthemke and R. Lefever, \textit{Noise-induced transitions:
Theory and applications in physics, chemistry, and biology}
(Springer-Verlag, Berlin and New York, 1984).

\bibitem{wei} U. Weiss, \textit{Quantum Dissipative systems} (World Scientific,
Singapore, 1999).

\bibitem{lui} W. H. Louisell, \textit{Quantum Statistical Properties of
Radiation} (J. Wiley, 1973).

\bibitem{bao} J. D. Bao, Phys. Rev. A {\bf 65}, 052120-052127
(2002).

\bibitem{bao1} J. D. Bao, Phys. Rev. A {\bf 69}, 022102-022106 (2004).

\bibitem{tani} K. Okamura and Y. Tanimura, Phys. Rev. E {\bf 56},
2747-2750; T. Steffen and Y. Tanimura, J. Phys. Soc. Jap. {\bf
69}, 3115-3132; T. Kato and Y. Tanimura, J. Chem. Phys. {\bf 117}
6221-6234 (2002); \textit{ibid.} {\bf 120}, 260-271 (2004).

\bibitem{hil} E. P. Wigner, Phys. Rev. {\bf 40}, 749-759 (1932);
M. Hillery, R. F. O'Connell, M. O. Scully and E. P. Wigner, Phys.
Rep. {\bf 106}, 121-167 (1984).

\bibitem{zwa} R. Zwanzig, Lectures in Theoretical Physics, Vol.3;
J. Stat. Phys. {\bf 9}, 215-220 (1973).

\bibitem{skb} S. K. Banik, B. C. Bag and D. S. Ray,
Phys. Rev. E {\bf 65}, 051106-051118 (2002).

\bibitem{db1} D. Banerjee, B. C. Bag, S. K. Banik and D. S. Ray,
Phys. Rev. E {\bf 65}, 021109-021121 (2002); D. Banerjee, S. K.
Banik, B. C. Bag, and D. S. Ray, Phys. Rev. E {\bf 66},
051105-051124 (2002).

\bibitem{db2} D. Banerjee, B. C. Bag, S. K. Banik and D. S. Ray, J. Chem.
Phys. {\bf 120}, 8960-8972 (2004)

\bibitem{bk1} D. Barik, S. K. Banik and D. S. Ray, J. Chem. Phys.
{\bf 119}, 680-695 (2003); D. Barik, B. C. Bag and D.S. Ray, J.
Chem. Phys. {\bf 119}, 12973-12980 (2003).

\bibitem{bk2} D. Barik and D. S. Ray, J. Chem. Phys. {\bf 121}, 1681-1689 (2004).

\bibitem{sm} B. Sundaram and P.W. Milonni,
Phys. Rev. E {\bf 51}, 1971-1982 (1995).

\bibitem{akp} A.K. Pattanayak and W.C. Schieve,
Phys. Rev. E {\bf 50}, 3601-3615 (1994).

\bibitem{gra} H. Grabert, P. Schramm and G. L. Ingold, Phys. Rep. {\bf 168},
115-207 (1988).

\end{thebibliography}
\end{document}